\documentclass[showpacs, pra,twocolumn,preprintnumbers ,amsmath, amssymb, superscriptaddress, aps]{revtex4-2}

\usepackage[utf8x]{inputenc}
\usepackage{color}
\usepackage{amsmath,amssymb}
\usepackage{pifont}
\usepackage{amssymb}  
\usepackage{bbold}
\usepackage{float}
\usepackage{subfloat}

\usepackage[caption=false]{subfig}
\usepackage{tikz}
\usepackage{makecell}
\usepackage{subfig}
\usepackage{pifont}   
\usepackage{graphicx} 
\graphicspath{{Figures2/}}
\usepackage{dcolumn}  
\usepackage{bm}       
\usepackage{multirow} 
\usepackage{placeins}
\usepackage[colorlinks]{hyperref}
\usepackage{mathtools}
\usepackage{appendix}

\def \be{\begin{align}}
	\def \ee{\end{align}}
\def \bea{\begin{eqnarray}}
	\def \eea{\end{eqnarray}}

\begin{document}
	
	\title{Confinement in a magnetically induced WSe$_2$ quantum dots}
		\author{Rachid El Aitouni}
		\affiliation{Laboratory of Theoretical Physics, Faculty of Sciences, Choua\"ib Doukkali University, PO Box 20, 24000 El Jadida, Morocco}
\author{Mohammed El Azar}
\affiliation{Laboratory of Theoretical Physics, Faculty of Sciences, Choua\"ib Doukkali University, PO Box 20, 24000 El Jadida, Morocco}

\author{Clarence Cortes}
	\affiliation{Vicerrector\'ia de Investigaci\'on y Postgrado, Universidad de La Serena, La Serena 1700000, Chile}

    	\author{David Laroze}
	\affiliation{Instituto de Alta Investigación, Universidad de Tarapacá, Casilla 7D, Arica, Chile}

	\author{Ahmed Jellal}
	\email{a.jellal@ucd.ac.ma}
	\affiliation{Laboratory of Theoretical Physics, Faculty of Sciences, Choua\"ib Doukkali University, PO Box 20, 24000 El Jadida, Morocco}

\begin{abstract}

Monolayer tungsten diselenide (WSe$_2$) has become a suitable platform for quantum transport and spintronics and valleytronics applications because it possesses an intrinsic band gap and strong spin-orbit coupling and spin-valley coupling features. The electrostatic confinement of Dirac fermions proves challenging in graphene because of Klein tunneling yet WSe$_2$ provides an environment that supports both carrier localization and the development of confined quantum states. 
In this work, we theoretically investigate the confinement of massive Dirac fermions in a WSe$_2$ quantum dot generated by a localized magnetic field. Using the effective Dirac Hamiltonian in the presence of a magnetic flux, we derive the exact wave functions and scattering coefficients by employing Kummer’s confluent hypergeometric functions together with Bessel and Hankel functions. Our results show that the localized magnetic field provides an efficient mechanism to suppress Klein tunneling and promote the formation of stable quasibound states. We systematically examine the scattering efficiency and carrier density distributions as functions of the incident energy, magnetic field strength, and quantum dot radius. We find that low-energy carriers are strongly confined by the magnetic barrier, while the interplay between magnetic localization and geometric confinement gives rise to sharp and tunable resonance peaks. These results provide valuable insight into the control of spin--valley transport in transition metal dichalcogenide nanostructures and establish a theoretical basis for the development of quantum confinement devices and quantum information technologies.

\end{abstract}
		
	\maketitle
	\section{Introduction}

The outstanding achievements in two-dimensional (2D) materials research have opened new possibilities for developing nanoelectronics and spintronics and valleytronics applications \cite{Novoselov2016,Xu2014,Schaibley2016}. After graphene established its initial presence in this research area, semiconducting transition metal dichalcogenides (TMDs) became important materials because their natural band gap properties and strong spin--orbit coupling and their ability to connect spin with valley degrees of freedom \cite{Mak2010,Xiao2012}. Monolayer tungsten diselenide (WSe$_2$) has become the focus of research because it serves as an effective platform to create confined quantum states and develop adjustable nanoscale devices \cite{Koperski2015,Srivastava2015,Fang2012}.
WSe$_2$ demonstrates a direct band gap which exists at both the $K$ and $K'$ valleys of its Brillouin zone \cite{Katsnelson2006,Zhao2013,Kormanyos2015} while graphene presents difficulties with electrostatic confinement through Klein tunneling because its Dirac spectrum contains no energy gap. The main property of this material establishes a framework for carrier movement while creating new possibilities to study different types of material confinement effects \cite{Kormanyos2014,Klinovaja2013,Wang2018}. WSe2 demonstrates a strong intrinsic spin--orbit interaction which generates substantial electronic band spin splitting thus enabling researchers to study how magnetic confinement interacts with spin effects and valley-dependent transport properties \cite{Zhu2011,Liu2013,Aivazian2015}.

Quantum dots (QDs) represent an important platform for studying carrier confinement in low-dimensional systems \cite{Loss1998,Hanson2007,Kouwenhoven2001}. In 2D materials, QDs enable the exploration of discrete energy spectra, resonant scattering phenomena, and localized charge density distributions, which are of great interest for both fundamental research and potential device applications \cite{Palacios2017,Recher2010}.
{Although conventional confinement strategies predominantly rely on electrostatic gating, the implementation of spatially inhomogeneous magnetic profiles to engineer magnetic quantum dots presents a highly compelling alternative paradigm for the localization of charge carriers \cite{DeMartino2007, Matulis1994}.}
{These magnetically engineered architectures are of profound interest when applied to Dirac-type frameworks. In such regimes, the intricate interplay among the geometric boundary conditions, the orbital angular momentum, and the applied magnetic flux facilitates the emergence of long-lived quasi-bound states and highly prominent resonant scattering signatures \cite{DellAnna2009,Masir2008,penna2022,elazarterm2024,elazarflux2024}.}

For monolayer WSe$_2$, the study of magnetically induced quantum dots is of special interest because the intrinsic gap and spin-orbit coupling can strongly modify the confinement mechanism compared with graphene-based systems \cite{Katsnelson2006,Xiao2012}. 
{The application of a spatially restricted magnetic profile is anticipated to go beyond merely altering the orbital dynamics of the carriers; it fundamentally dictates the distribution of available scattering channels, the resonant characteristics of the scattering cross-section, and the precise spatial mapping of the electron probability density both within and adjacent to the quantum dot. A rigorous theoretical elucidation of these underlying mechanisms is therefore imperative to fully assess the technological viability of WSe$_2$-based nanoscale architectures in the realms of quantum transport and advanced confinement engineering \cite{Beenakker2008,Park2008,Bouhlallaser2025}.}

Magnetic quantum dots have been extensively studied in the context of 
graphene~\cite{Matulis2008,Masir2009}, where the linear 
gapless Dirac spectrum and the absence of Klein-tunneling suppression 
make electrostatic confinement problematic. In those works, a spatially 
restricted magnetic field $B(r)=B\,\Theta(R-r)$ was shown to produce 
quasibound states and resonant scattering through the interplay of 
cyclotron motion and geometric boundary effects. The model we present 
here shares the same magnetic-confinement geometry, but the physics is 
fundamentally altered by three material-specific features of monolayer 
WSe$_2$. ({\it i}) A finite direct band gap $\Delta\approx 0.85$~eV, 
which strongly suppresses Klein tunneling and enables much more 
efficient carrier localization than in graphene. ({\it ii}) Strong 
intrinsic spin--orbit coupling ($\lambda_c$, $\lambda_v$), which lifts 
the spin and valley degeneracies and creates spin-split resonant 
channels.  ({\it iii}) Spin-valley locking, which ties the 
magnetic confinement to valley-dependent transport. Consequently, 
whereas graphene magnetic dots are governed by a single valley and 
spinless Dirac cone, the WSe$_2$ dot exhibits valley-contrasting 
scattering efficiencies and spin-resolved quasibound states that have 
no counterpart in gapless systems. Our results therefore extend the 
magnetic-dot paradigm to a massive, spin--orbit-coupled Dirac 
material and establish a theoretical basis for spintronic and 
valleytronic confinement devices.

We study the confinement of massive Dirac fermions in a monolayer WSe$_2$ sheet subjected to a magnetic field localized within a circular region, forming a magnetic quantum dot. By employing an effective Dirac-like Hamiltonian that incorporates the intrinsic band gap, strong spin--orbit coupling, and spin--valley interactions, we derive the exact wave functions both inside and outside the dot in the presence of a magnetic field. The analytical solutions are obtained using Kummer’s confluent hypergeometric functions inside the dot and Bessel/Hankel functions in the outer region. By imposing the continuity of the spinor wave functions at the interface, we determine the corresponding scattering amplitudes and reflection coefficients. We then perform a systematic analysis of the scattering efficiency and the electronic density distributions as functions of the incident energy, magnetic field strength, and quantum dot radius. Our results show that the localized magnetic field provides an efficient mechanism to suppress Klein tunneling and enhance carrier localization, leading to the formation of stable quasibound states. In particular, low-energy electrons are strongly trapped by the magnetic barrier, while the interplay between magnetic confinement and geometric confinement gives rise to sharp and tunable resonance peaks. These findings demonstrate the feasibility of magnetic confinement in WSe$_2$-based quantum dots and provide useful insight for the development of spintronic, valleytronic, and quantum information devices based on transition metal dichalcogenides.

{The structure of the present paper  is as follows. Sec. \ref{TM} describes the theoretical model and mathematical framework used to solve the wave equations in the presence of a magnetic field. Sec. \ref{SSPP} is devoted to the scattering properties, where we explicitly determine the 
 reflected current density and efficiency. In Sec. \ref{Num}, we numerically analyze our results, with an emphasis on the emergence of quasi-bound states, the behavior of the scattering efficiency, and the spatial distribution of the electron density. Our main results  are summarized in Sec. \ref{con}, followed by  concluding remarks.}

		\section{Mathematical framework}\label{TM}

\begin{figure}[ht]
	\centering
    \includegraphics[scale=0.42]{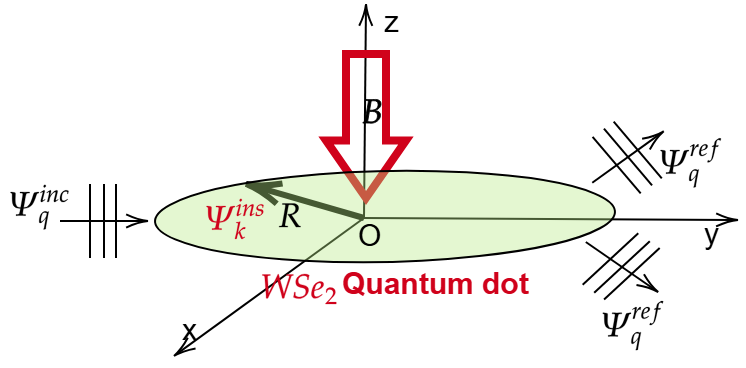}
	\caption{Schematic of a monolayer WSe$_2$ sheet lying in the $xy$-plane, 
	subjected to a perpendicular magnetic field $\vec{B}=B\hat{z}$ confined 
	to a circular region of radius $R$. The field is uniform inside the dot 
	($r<R$) and vanishes outside ($r>R$) in the symmetric gauge. Incident 
	Dirac waves $\Psi_q^{\mathrm{inc}}$ propagate along the $x$-direction, 
	penetrate the magnetic boundary, and form standing waves 
	$\Psi_k^{\mathrm{ins}}$ inside the dot, while reflected waves 
	$\Psi_q^{\mathrm{ref}}$ scatter outward. The inset coordinate system 
	shows the layer orientation with $\hat{z}$ normal to the WSe$_2$ plane.}
		 \label{str}
\end{figure}

We consider a monolayer WSe$_2$ sheet subjected to a magnetic field confined within 
a circular region of radius $R$, as illustrated in Fig. \ref{str}. The system is 
modeled as a quantum dot in which a uniform perpendicular magnetic field $B$ is 
present inside the dot region ($r < R$), while the exterior region ($r > R$) remains 
completely field-free. This configuration naturally divides the system into two 
distinct regions, each governed by the effective low-energy Dirac Hamiltonian 
appropriate to WSe$_2$, with the magnetic field entering through the minimal coupling 
of the vector potential in the interior region. The localized magnetic field preserves 
the rotational symmetry of the problem, allowing the total wave function to be 
decomposed into angular momentum channels and reducing the problem to a set of radial 
equations for each channel. 

The Hamiltonian describing the motion of an electron 
through the structure  can be written as
 \begin{equation}
 H = v_F \boldsymbol{\sigma} \cdot \left(\mathbf{p} + e\mathbf{A}\right)  -\lambda_c \tau s_z\frac{(\sigma_0+\sigma_z)}{2} - \lambda_v\tau s_z\frac{(\sigma_0-\sigma_z)}{2}
 \label{ham}
 \end{equation}
 where $v_F$ is the Fermi velocity, $\boldsymbol{\sigma} = (\sigma_x, \sigma_y)$ are 
 the Pauli matrices, $\mathbf{p} = -i\hbar\nabla$ is the momentum operator, $\mathbf{A}$  is the magnetic vector potential 
$\lambda_c$ and $\lambda_v$ denote the spin-orbit coupling parameters for conductance and valence bands, with $\tau=1$ (-1) for valley $K$($K^\prime$) and $s_z=1$(-1) spin up(down), ${\sigma_0}$ is the $2\times 2$ identity matrix in the pseudospin space.
In writing the vector potential as $\vec{A}=(Br/2)\,\vec{e}_\varphi$ 
for $r<R$ and $\vec{A}=0$ for $r>R$, one must be aware that $A_\varphi$ 
is discontinuous at $r=R$. Since the physical magnetic field is 
$B_z=(1/r)\,\partial_r(rA_\varphi)$, this discontinuity produces a 
delta-function contribution
\begin{equation}
	B_z(r)=B\,\Theta(R-r)-\frac{BR}{2}\,\delta(r-R)
\end{equation}
namely a ring-like magnetic field localized at the boundary that exactly 
compensates the interior flux, so that the net magnetic flux vanishes. 
An alternative, commonly adopted in the magnetic quantum-dot 
literature~\cite{Matulis2008,Masir2009}, is the continuous vector 
potential $A_\varphi=BR^2/(2r)$ for $r>R$. That choice eliminates the 
boundary singularity but introduces a long-range $1/r$ vector potential 
outside the dot (equivalent to an Aharonov-Bohm flux tube of total flux 
$\pi B R^2$), thereby changing the asymptotic scattering states. In the 
present work we retain the discontinuous form, which corresponds to a 
magnetic dot compensated by an opposite boundary ring, so that the net 
flux is zero.

Because the system possesses circular symmetry, we work in polar coordinates 
and write the Hamiltonian as
 \begin{widetext}
     \begin{align}
	H =  v_F \left( \tau \sigma_r p_r + \sigma_\varphi (p_\varphi-eA_B \right) + \Delta\sigma_z -\lambda_c \tau s_z\frac{(\sigma_0+\sigma_z)}{2} - \lambda_v\tau s_z\frac{(\sigma_0-\sigma_z)}{2}.
\end{align}
 \end{widetext}
Using the matrix representation of the Pauli matrices in polar coordinates,
\begin{equation}
\sigma_r =
\begin{pmatrix}
0 & e^{-i\varphi} \\
e^{i\varphi} & 0
\end{pmatrix},
\qquad
\sigma_\varphi =
\begin{pmatrix}
0 & -i e^{-i\varphi} \\
i e^{i\varphi} & 0
\end{pmatrix}
\end{equation}
the Hamiltonian can be projected onto the pseudospin basis
$(\psi_A,\psi_B)^T$. The terms proportional to $\sigma_z$ and   the projectors $(\sigma_0\pm\sigma_z)/2$ contribute only to the diagonal elements of the Hamiltonian, whereas the radial and angular kinetic operators couple the two spinor components. This leads to the following matrix form of the
Hamiltonian
\begin{widetext}
    \begin{align}
	 H = \begin{pmatrix} \Delta - \lambda_c \tau s_z &v_F e^{i\varphi}  \left( \tau p_r - i \big(p_\varphi + eA_B\big) \right) \\ v_F e^{i\varphi}\left( \tau p_\varphi + i \big(p_\varphi + eA_B\big) \right) & -\Delta - \lambda_v \,\tau s_z \end{pmatrix}\label{H1}.
\end{align}
\end{widetext}

{Since the Hamiltonian $H$ commutes with the total angular momentum operator $J_z$, the variables can be separated. Consequently, the spinor  solution of the eigenvalue equation $H\Psi=E\Psi$ can be expressed as}
\begin{align}
		\Psi_k(r,\varphi) =
	\begin{pmatrix}
		\psi^A_k(r)\, e^{i l \varphi} \\
		i\,\psi^B_k(r)\, e^{i (l+1)\varphi}
	\end{pmatrix}
\end{align}
{where $l$ is the angular momentum quantum number.} The eigenvalue equations are given by
	\begin{align}
		E_c\psi^A_k(r)
		&= \hbar v_F \left[
		-  \tau \frac{d}{dr}
		-  \left( \frac{ l+1}{r} - \frac{r}{2\ell^2_B } \right)
		\right] \psi^B_k(r)\label{valeq1}
		\\
	E_v\psi^B_k(r)
		&= \hbar  v_F\left[
		-  \tau  \frac{d}{dr}
		+  \left( \frac{ l}{r} -  \frac{r}{2\ell^2_B } \right)
		\right] \psi^A_k(r)\label{valeq2}
	\end{align}
		with $E_c = E-\Delta +\lambda_c \tau s_z$, 
	$E_v =E+\Delta+\lambda_v \tau s_z$, and
	$l_B = \sqrt{\frac{\hbar}{e B}}$. Form the two equations, we can obtain the homogene equation from the two componenets ($\tau=1$)
  \begin{align}\label{7777}
\left[ \frac{d^{2}}{dr^{2}} +\frac{1}{r}\frac{d}{dr} + k^2+ \frac{l+1}{\ell_B^{2}}-\frac{l^2}{r^{2}} - \frac{r^{2}}{4\,\ell_B^{4}} \right]\psi^A_k(r) = 0
 \end{align}
where  $k^2=\frac{E_c*E_v}{\hbar^2 v_F^{2}} $, thne we can substitution $z = \frac{r^{2}}{2\ell_B^{2}}$, and $\psi^B_k(r) = r^{|l|} e^{-\tfrac{r^{2}}{4\ell_B^{2}}} f(z)$, the equation becomes the confluent hypergeometric equation (Kummer’s equation) : 
\begin{equation} 
	z \frac{d^{2}f(z)}{dz^{2}} + (b - z)\frac{df(z)}{dz} - a f(z) = 0
	 \end{equation} 
	 	 with $b = |l| + 1$ and $a =\frac{|l|-l}{2} - \frac{\ell_B^{2}}{2}k^2$.
     The general solution is given by
   \begin{equation} 
   	\psi^A_k(r) = r^{|l|} e^{-\frac{r^{2}}{4\ell_B^{2}}} \left[ c_1 M\left(a,b,\frac{r^{2}}{2\ell_B^{2}}\right) + c_2 U\left(a,b,\frac{r^{2}}{2\ell_B^{2}}\right) \right]
    \end{equation}
 with $c_1$, $c_2$ are arbitrary constants and $M(a,b,z)$ et $U(a,b,z)$ are the independent solutions of Kummer’s equation written as 
 \begin{widetext}
     \begin{align}
    	&M(a,b,c) \equiv {}_1F_1(a; b; c) = \sum_{n=0}^{\infty} \frac{(a)_n}{(b)_n} \frac{(c)^n}{n!}\\
    	&U(a,b,x) = \frac{\pi}{\sin(\pi b)} \left[ \frac{M(a,b,c)}{\Gamma(1+a-b)} - (c)^{1-b}\,\frac{M(1+a-b,2-b,c)}{\Gamma(a)} \right]
    \end{align} 
 \end{widetext} 
where $_1F_1(a; b; x)$ Hypergeometric function, $(a)_n = a(a+1)\cdots(a+n-1) = \frac{\Gamma(a+n)}{\Gamma(a)}$ is the Pochhammer symbol, and $(b)_n = \frac{\Gamma(b+n)}{\Gamma(b)}$ ($\Gamma(n)=(n-1)!$).
To obtain physically acceptable solutions, it is necessary that they converge as $r\to 0$ ($l\neq 0$), then $\psi^B_k(r) \sim c_2  r^{|l|} \left( \frac{r^{2}}{2\ell_B^{2}} \right)^{1-(|l|+1)} \sim c_2  r^{-|l|} \to \infty$ which implies that $c_2=0$. Finally, the general solution is written as follows:
	\begin{widetext}
	\begin{align} 
	&	\psi^{A,+}_k = c^l_t r^{l} e^{-\frac{r^{2}}{4\ell_B^{2}}}   \,_1F_1\left(- \frac{\ell_B^{2}}{2}k^2,1+l,\frac{r^{2}}{2\ell_B^{2}}	\right), \quad l>0\\
	&\psi^{A,-}_k = c^l_t r^{-l} e^{-\frac{r^{2}}{4\ell_B^{2}}}   \,_1F_1\left(-l- \frac{\ell_B^{2}}{2}k^2,1-l,\frac{r^{2}}{2\ell_B^{2}}	\right),\quad l<0
		\end{align}
	\end{widetext}
In the same way, one can find the other component $\psi^{A}_k(r)$, either by substituting into the eigenvalue equation (\ref{valeq1},\ref{valeq2}), which gives
\begin{widetext}
    \begin{align} 
	&\psi^{B,+}_k = c^l_t\frac{ k}{2}r^{1+l} e^{-\tfrac{r^2}{4\ell_B^2}}\,_1F_1\left(1 - \frac{\ell_B^{2}}{2}k^2, l+ 2, \tfrac{r^2}{2\ell_B^2}\right),\quad l>0\\
	&\psi^{B,-}_k = c^l_tr^{1+l} e^{-\tfrac{r^2}{4\ell_B^2}}\left[ \frac{2l}{k}{}_1F_1\left(-l- \frac{\ell_B^{2}}{2}k^2,  1-l, \tfrac{r^2}{2\ell_B^2}\right) + \frac{(2l+\ell_B^2 k^2)r^2}{2 \ell_B^2 k (1+l)}\,_1F_1\left(1-l- \frac{\ell_B^{2}}{2}k^2, 2-l, \tfrac{r^2}{\ell_B^2}\right) \right],\quad l<0
	\end{align}
\end{widetext}
	Finally, the wave function inside the quantum dots is
    \begin{widetext}
        \begin{align}
	\Psi^{ins}_k(r,\varphi) =\sum_{l=-\infty}^{-1}
	\begin{pmatrix}
		\psi^{A,-}_k(r)\, e^{i l \varphi} \\
		i\,\psi^{B,-}_k(r)\, e^{i (l+1)\varphi}
	\end{pmatrix}+\sum_{l=0}^{+\infty}\begin{pmatrix}
	\psi^{A,+}_k(r)\, e^{i l \varphi} \\
	i\,\psi^{B,+}_k(r)\, e^{i (l+1)\varphi}
	\end{pmatrix}.
\end{align}
    \end{widetext}
In order to obtain the wave functions outside the quantum dot, we proceed by solving the corresponding eigenvalue equation. In this case, equation \eqref{7777} reduces to
\begin{equation}
	\left[\frac{d^{2}}{dr^{2}}
	+ \frac{1}{r}\,\frac{d}{dr}
	+ 
	k^2 - \frac{l(l+1)}{r^{2}}
	\right]\psi^{A}_k(r) = 0
\end{equation}
which can be solved to obatin
\begin{align} \Psi^\text{out}_q(r,\varphi) = \Psi^\text{inc}_q(r,\varphi)+\Psi^\text{ref}_q(r,\varphi)
  \end{align}
where the incident and reflected parts are given by
  \begin{align}
  	&\Psi^\text{inc}_q(r,\varphi)=\sum_{l=-\infty}^{\infty}i^l\begin{pmatrix}J_l(q r)e^{il\varphi}\\iJ_{l+1}(q r)e^{i(l+1)\varphi}(q r) \end{pmatrix}\\
  	& \Psi^\text{ref}_q(r,\varphi)=\sum_{l=-\infty}^{\infty}c_r^l \begin{pmatrix}H^{(1)}_l(q r)e^{il\varphi}\\i^lH^{(1)}_{l+1}(q r)e^{i(l+1)\varphi} \end{pmatrix}
  \end{align}
where $H^{(1)}_n(x)$ are the first ordre Hankel fonctions.

\section{Scattering properties}\label{SSPP}
Having obtained the energy spectrum and the corresponding wave functions in both 
regions, we now turn to the scattering properties of the WSe$_2$ quantum dot. 
The analytically derived solutions serve as the fundamental building blocks for 
constructing the scattering matrix, from which the scattering coefficients are 
extracted by imposing the continuity of the wave functions at the boundary $r = R$. 
These coefficients fully characterize how incident electrons interact with the 
magnetic quantum dot, and allow us to compute key scattering observables such as 
the scattering efficiency and the spatial density profiles. Through these quantities, 
we can systematically investigate the conditions under which quasibound states emerge 
as sharp resonances, and assess the role of the magnetic field strength, dot radius, 
and incident energy in controlling carrier localization within the dot. As a result, we have
\begin{align}
	\Psi_q^\text{inc}(r=R,\varphi)+\Psi_q^\text{ref}(r=R,\varphi)=\Psi_k^\text{ins}(r=R,\varphi)
\end{align}
and after some algebra, we find the coefficients $c_r^l$ and $c_t^l$ 
\begin{align}
	& c_t^l 
        =\frac{\sqrt{2}e^{i\frac{(l+1)\pi}{2}}} {\pi\,q R\left[H^{(1)}_{l}(q R)\,\psi^{B,\pm}_k(R)	- \,H^{(1)}_{l+1}(q R)\,\psi^{A,\pm}_k(R)\right]}
\\
&c_r^l
=\frac{J_{l}(q R)\,\psi^{B,\pm}_k(R) - J_{l+1}(q R)\,\psi^{A,\pm}_k(R)} {H^{(1)}_{l}(q R)\,\psi^{B,\pm}_k(R) - H^{(1)}_{l+1}(q R)\,\psi^{A,\pm}_k(R)}.
 \end{align}

{To characterize the scattering process, we consider stationary states of fixed energy. Consequently, the probability density $\rho(r,\varphi) = \Psi^\dagger(r,\varphi)\Psi(r,\varphi)$ is time-independent. The associated probability current density for the Dirac spinor is defined as $\mathbf{J}(r,\varphi) = v_F\Psi^\dagger(r,\varphi)\boldsymbol{\sigma}\Psi(r,\varphi)$. Under these steady-state conditions, the local continuity equation simplifies to the spatial conservation of the probability 
\begin{equation}
\nabla\cdot\mathbf{J}(r,\varphi) = 0.
\label{cont}
\end{equation}
In our scattering geometry, the total wave function outside the quantum dot is a superposition of incident and reflected waves, while the inner region is governed by the transmitted solution. Consequently, the relevant physical quantity for determining the scattering cross-section is the radial component of the reflected current. This quantity is obtained by projecting the reflected current density along the radial direction
\begin{equation}
J^\text{ref}_{r}(r,\varphi) = v_F \left(\Psi^\text{ref}\right)^\dagger \sigma_r \Psi^\text{ref}.
\end{equation}
From the current densities satisfying the continuity equation (\ref{cont}), and taking into account the asymptotic behavior of the first-order Hankel function in the far-field limit ($qr \gg 1$), namely
\begin{equation}
H_l^{(1)}(qr) \simeq \sqrt{\frac{2}{\pi qr}} \exp\left[i\left(qr-\frac{l\pi}{2}-\frac{\pi}{4}\right)\right]
\end{equation}
one obtains the explicit far-field expression for the reflected radial current
\begin{align}
J^\text{ref}_{r}(\varphi) =& \frac{4}{\pi q r} \sum_{l=-\infty}^{+\infty} |c_l^r|^2 \\
&+ \frac{8}{\pi q r} \sum_{l<l'} \mathrm{Re} \left[ c_l^r\left(c_{l'}^r\right)^* \right] \cos[(l-l')\varphi]
\label{Jref}\notag
\end{align}
where the first term   represents the incoherent contribution of each partial wave, whereas the second term encapsulates the quantum interference between different angular momentum channels. When calculating the total reflected flux by integrating over a circle of large radius $r$, these interference terms naturally vanish due to the angular orthogonality condition
\begin{equation}
\int_0^{2\pi}\cos[(l-l')\varphi]\,d\varphi = 0,  \quad l \neq l'.
\end{equation}
Thus, the total reflected flux simply reduces to the sum of the individual channel contributions
\begin{equation}
I^\text{ref} = \int_0^{2\pi}J^{ref}_{r}(\varphi)\,r\,d\varphi = \frac{8}{q} \sum_{l=-\infty}^{+\infty} |c_l^r|^2.
\end{equation}

The scattering cross-section $\sigma$ is fundamentally defined as the ratio between the total reflected flux and the incident flux per unit length. Assuming a normalized incident Dirac plane wave (where the incident flux is taken as unity), the scattering cross-section is directly given by:
\begin{equation}
\sigma = \frac{I^\text{ref}}{I^\text{inc}} = \frac{8}{q} \sum_{l=-\infty}^{+\infty} |c_l^r|^2.
\end{equation}

Finally, to provide a dimensionless measure of the scattering strength, the scattering efficiency $Q$ is defined by normalizing the scattering cross-section by the geometrical cross-section of the circular quantum dot \cite{Heinisch2013,Schulz2015}. Hence, we obtain the following expression
\begin{equation}
Q = \frac{\sigma}{2R} = \frac{4}{qR} \sum_{l=-\infty}^{+\infty} |c_l^r|^2.
\label{Q}
\end{equation}
 }
 \section{Numerical Results}\label{Num}
 {Having analytically established the fundamental solutions and wave functions for our system, we now proceed to a systematic numerical investigation of the 
scattering processes within the WSe$_2$ magnetic quantum dot. Our study focuses on the numerical evaluation of the scattering efficiency $Q$ and the spatial mapping of the electronic density $\rho(r,\varphi)$ in and around the magnetic confinement region. We pay particular attention to the interplay 
between the sizable intrinsic bandgap and the robust spin-orbit coupling, examining how these characteristic features of monolayer WSe$_2$ modulate the resonant spectra and the spatial localization of charge carriers.}

\begin{figure*}[t]
	\centering
	{\subfloat[$E=0.8$ eV]{\includegraphics[scale=0.35]{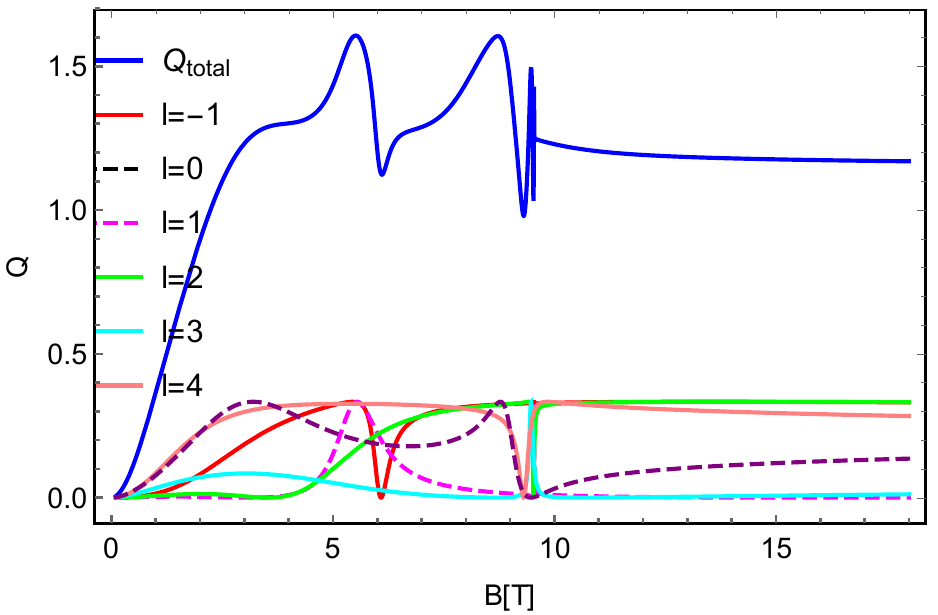}}\label{QE095up}}
    {\subfloat[$E=1$ eV]{\includegraphics[scale=0.35]{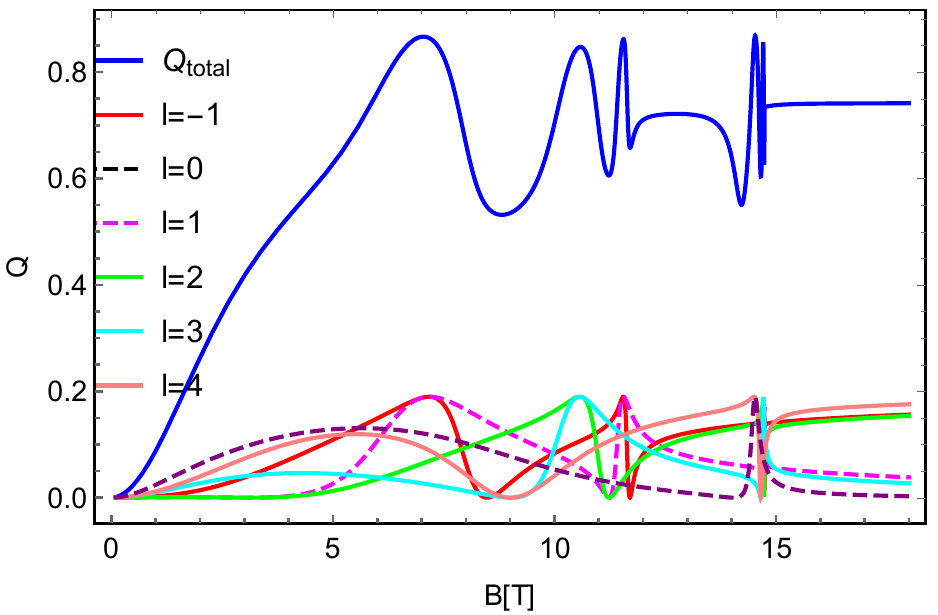}}\label{QE1up}}
    {\subfloat[$E=2$ eV]{\includegraphics[scale=0.35]{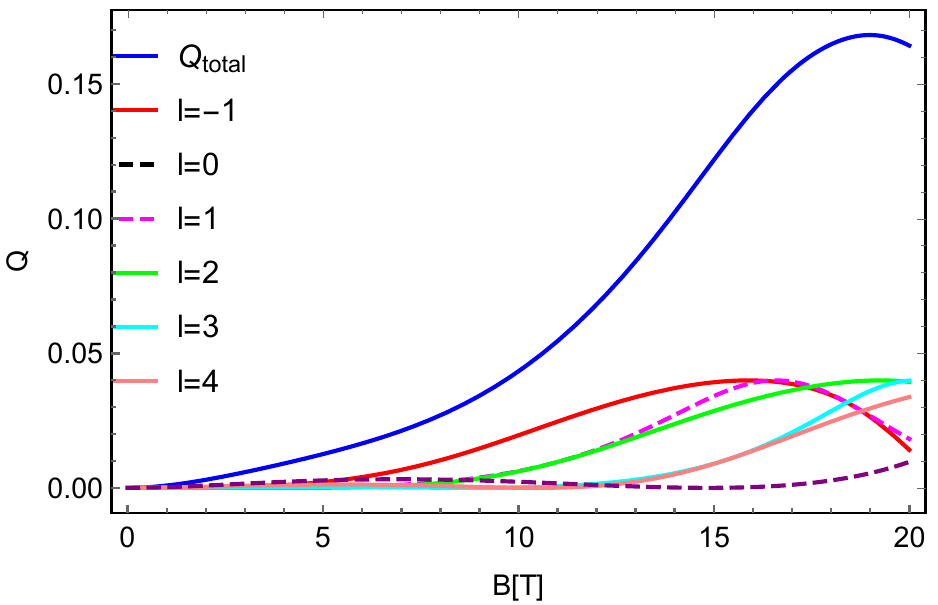}}\label{QEB2}}
	\caption{(Color online) The scattering efficiency $Q$ as a function of the magnetic field $B$ for different values $E$ ($0.8, 1, 2$ eV), for $R=50$ nm with intrinsic values $\lambda_c=75$ meV, $\lambda_v=112,5$ meV, $\Delta=0.85$ eV.}\label{QB}
\end{figure*}

\begin{figure*}[t]
	\centering
    {\subfloat[$B=1$ T]{\includegraphics[scale=0.35]{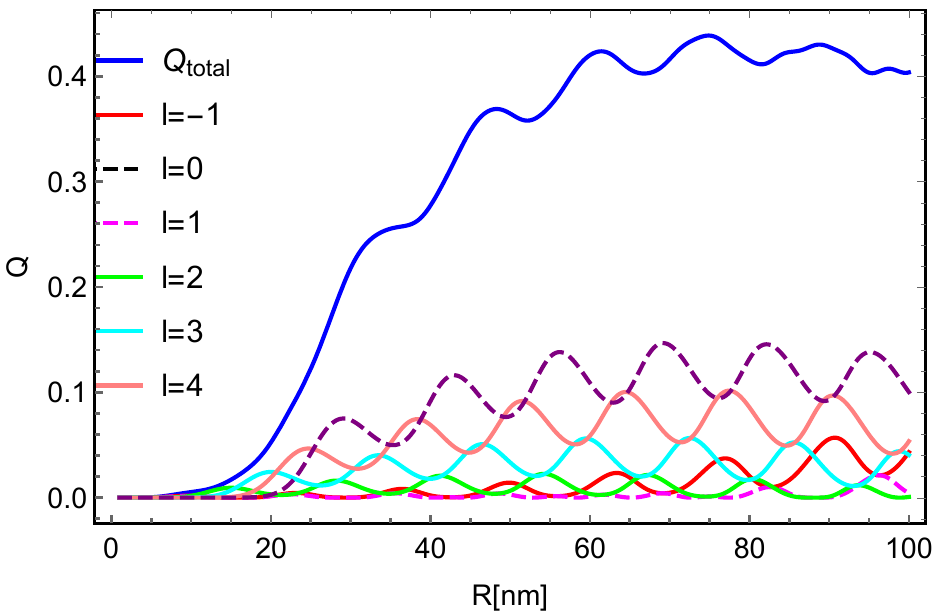}}}\label{QRB1}
    {\subfloat[$B=5$ T]{\includegraphics[scale=0.35]{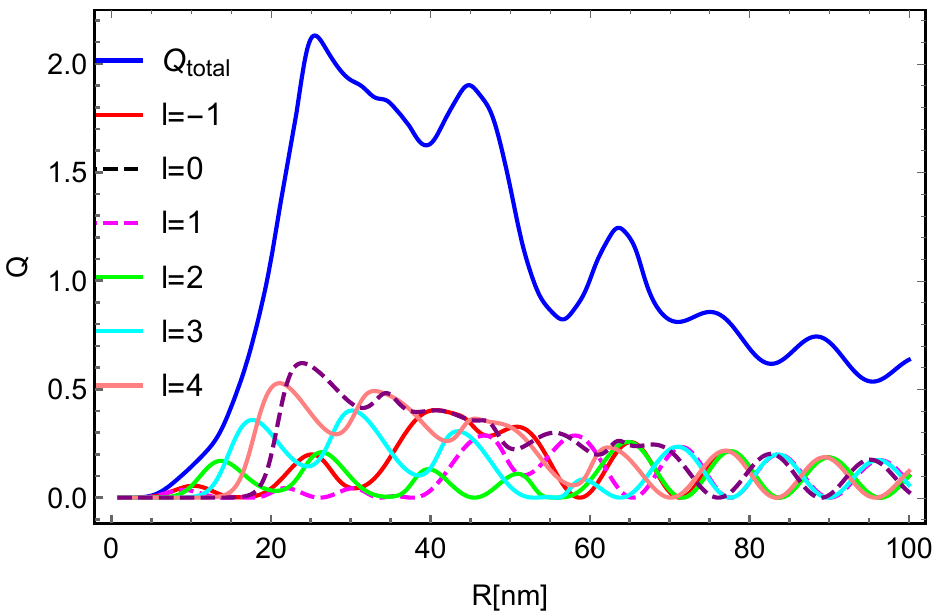}}}\label{QRB5}
    {\subfloat[$B=10$ T]{\includegraphics[scale=0.35]{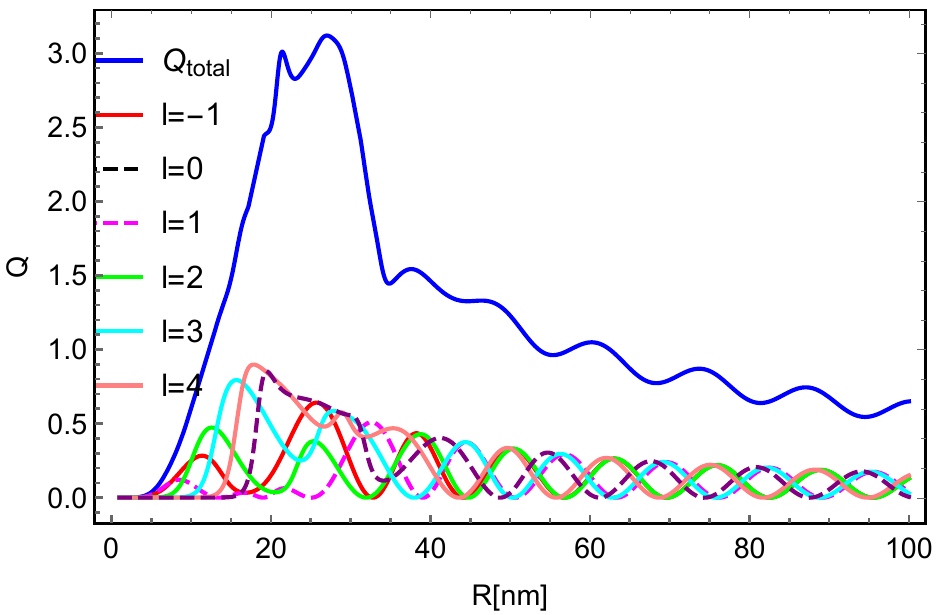}}}\label{QRB10}
	\caption{(Color online) The scattering efficiency $Q$ as a function of the radius for three values of magnetic field ($B=1,5,10$ T), with $E=1$ eV.}\label{QRB}
\end{figure*}

Figure~\ref{QB} shows the total scattering efficiency $Q_{total}$ along with the partial-wave contributions for various angular momentum channels $l=-1,0,1,2,3,4$, charted as a function of the perpendicular magnetic field $B$ for three incident energies $E=0.8$, $1$, and $2$ eV, while keeping the quantum dot radius fixed at $R=50$ nm.
The scattering efficiency experiences a significant decrease when the incident energy reaches higher levels. The total scattering efficiency for $E=0.8$ eV shows a high value which increases with magnetic field strength because the dot exhibits strong carrier confinement properties. High-energy carriers show reduced sensitivity to both the confining potential and the magnetic field at $E=2$ eV which results in minimal total scattering. The same pattern at $E=1$ eV shows a reduced strength compared to the initial results.
Low-energy Dirac carriers have longer effective interaction periods with the confined area, therefore increasing resonant scattering, hence explaining this conduct. High-energy electrons, by contrast, have greater kinetic energy and cross the QD area more readily, therefore lowering the likelihood of reflection and localization. Similar energy-dependent suppression of scattering has also been reported for graphene quantum dots and magnetic barriers, where increasing carrier energy weakens quasibound-state formation and reduces confinement effects \cite{Matulis2008,Ramezani2010}.
By generating cyclotron motion and magnetic localization, the magnetic field improves confinement. The cyclotron radius lowers as $B$ rises, hence compelling the electronic pathways to curve more sharply around the quantum dot area. This raises the overall scattering efficiency by thereby increasing the backscattering likelihood. Particularly for the lowest angular momentum channels, especially $l=0$ and $l=1$, which control transport at low energies, the impact is quite strong. In graphene magnetic dots, where magnetic localization rather than electrostatic trapping causes confinement, Matulis and Peeters discussed a similar process \cite{Matulis2008}.
Another interesting feature is the uneven contribution of the angular momentum modes. While channels with higher $l$ stay suppressed, lower-order modes contribute substantially more than their higher-order counterparts. This shows how magnetic confinement prefers low angular momentum resonances while higher-order modes demand more kinetic energy to actively participate.
Qualitatively distinct from graphene quantum dots, where the isotropic Dirac cone creates a more symmetrical angular distribution of scattering channels \cite{ Matulis2008, Ramezani2010}. Magnetic confinement mostly changes Landau quantization and suppresses Klein tunneling in graphene. However, in WSe$_2$, the intrinsic band gap, intense spin--orbit coupling, and valley-dependent mass terms produce stronger localization and greater magnetic sensitivity \cite{Xiao2012}. Coexistence of spin-valley locking and large Dirac fermions in monolayer transition-metal dichalcogenides like WSe$_2$ strengthens magnetic confinement's influence compared to gapless graphene systems \cite{Xiao2012}. Consequently, in WSe$_2$-based quantum dots, the growth of scattering efficiency with $B$ is noticeably more significant.

\begin{figure*}[t]
	\centering
{\subfloat[ $B=1$ T]{\includegraphics[scale=0.45]{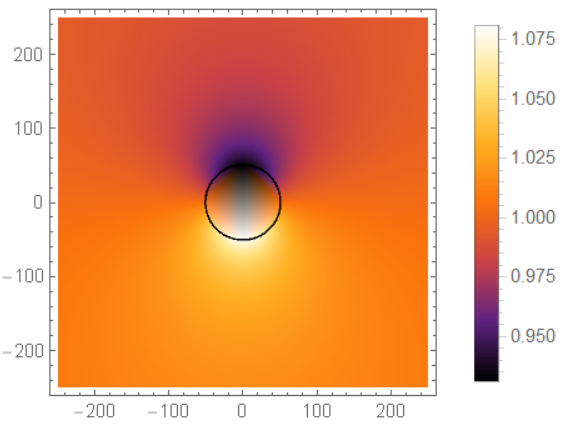}}}\label{DB1} {\subfloat[$B=5.6$ T]{\includegraphics[scale=0.45]{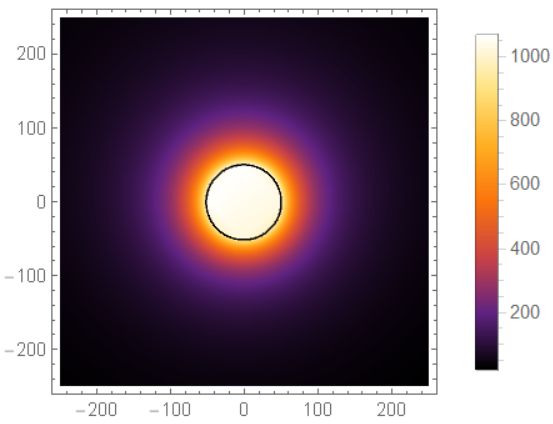}}}\label{DB5}
    {\subfloat[$B=9.26$ T]{\includegraphics[scale=0.45]{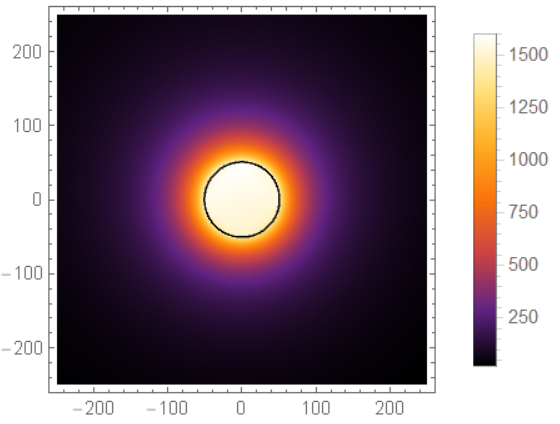}}}\label{DB9}\\
    {\subfloat[$B=9$ T]{\includegraphics[scale=0.45]{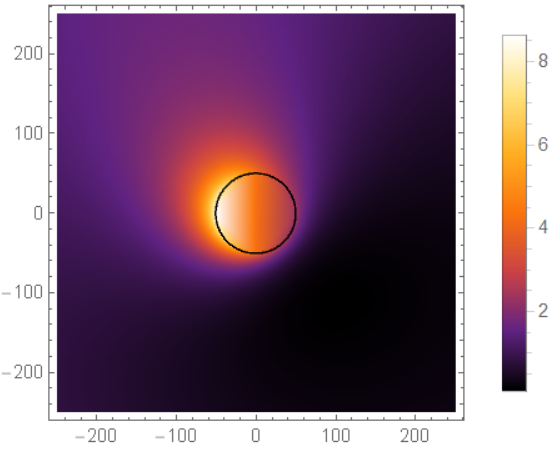}}}\label{DBB9}
    {\subfloat[ $B=7.02$ T]{\includegraphics[scale=0.45]{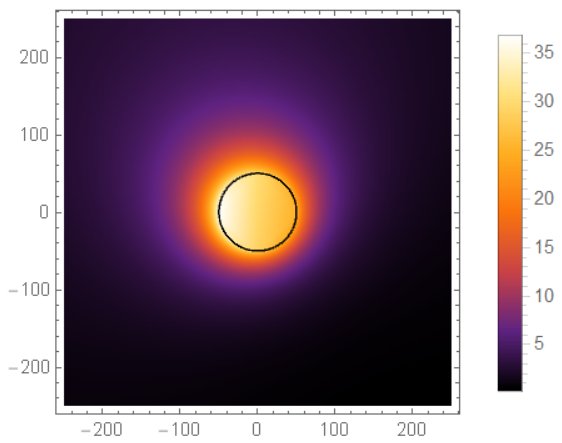}}}\label{DB1} {\subfloat[$B=11.6$ T]{\includegraphics[scale=0.45]{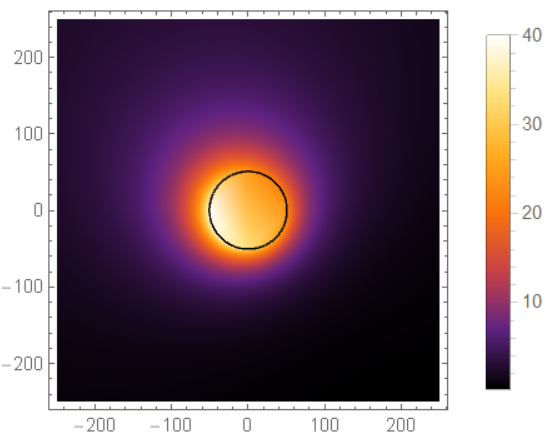}}}\label{DB5}
  	\caption{Density as a function of $x$ and $y$ for two energies: $E = 0.8$\, eV (top row) and $E = 1$\,eV (bottom row). The top row corresponds to magnetic fields $B = 1$\, T, $B = 5.9$\, T, and $B = 9.26$\, T, 
while the bottom row corresponds to $B = 9$, $B = 7.02$\, T, and $B = 11.6$\, T. These configurations illustrate the resonance intensities presented in Fig. \ref{QB}.}\label{RhoB}
\end{figure*}

Three values of the perpendicular magnetic field, $B=1$, 5, and 10 T, are shown in Figure~\ref{QRB}, along with the contributions of the several angular momentum channels $l=-1,0,1,2,3,4$ as functions of the quantum dot radius $R$ for the incident energy fixed at $E=1$ eV.
The results show beyond doubt that the quantum dot's radius influences its scattering efficiency. The overall scattering efficiency for small radii remains poor since the constrained area is too small to sustain strong resonant states or effective magnetic localization. Larger interaction region with a growing radius lets carriers spend more time inside the dot, then increasing the likelihood of scattering and temporary confinement. As a result, particularly for greater magnetic fields, $Q_{\text{total}}$ increases quickly with $R$.
The increase of $Q$ with $R$ at low magnetic field ($B=1$ T) is rather smooth and moderate, implying poor magnetic confinement. For $B=5$ T and particularly $B=10$ T, though, the scattering efficiency becomes significantly higher and more quickly with radius. This highlights the cooperative influence of geometric confinement and magnetic localization: a bigger radius gives a broader scattering region while a stronger magnetic field lowers the cyclotron radius and enhances carrier trapping inside the dot \cite{Matulis2008}.
Especially for medium radii, the partial-wave study also shows that the scattering mechanism is dominated by the lowest angular momentum channels. Most strongly contributing are the modes $l=0$ and $l=1$, since under cylindrical symmetry these channels match to the most efficient resonant states. Because they demand more orbital momentum and better coupling to be important, higher-order channels remain weaker. This conduct is in line with earlier investigations of magnetic quantum dots in graphene, whereby low-order angular momentum modes control quasibound-state creation \cite{Matulis2008,Ramezani2010}.
Because of its intrinsic band gap and strong spin--orbit coupling, WSe$_2$'s scattering efficiency is more dependent on the radius than graphene is. In graphene, Klein tunneling's presence and the lack of a band gap lower confinement efficiency even for rather huge dots \cite{ Matulis2008}. In WSe$_2$, however, the large Dirac spectrum reduces perfect transmission and boosts localization, therefore highlighting the radius dependence \cite{Xiao2012}. Further enhancing the transport's sensitivity to both the magnetic field and the dot size is the valley-contrasting spin splitting.
Hence, Figure~\ref{QRB} shows that in WSe$_2$, the quantum dot's radius acts as an effective control element for adjusting electronic scattering. Larger dots can greatly improve carrier confinement when coupled with a powerful magnetic field, which is much sought for uses in valley-selective transport systems, spin filters, and quantum confinement devices.

Figure~\ref{RhoB} depicts the probability density distribution as a function of the spatial coordinates $(x, y)$ for two incident energies, notably $E=0.8$ eV (top row) and $E=1$ eV (bottom row), under various values of the magnetic field $B$.   The top row represents $B=1$, $5.6$, and $9.26$ T, whereas the bottom row represents $B=9$, $7.02$, and $11.6$ T. These density maps directly relate to the resonance intensities seen in Figure~\ref{QB} and offer a visual representation of carrier localizing inside the magnetic quantum dot.
For the lower incident energy $E=0.8$ eV, the density profiles show great localization close the quantum dot area, especially as the magnetic field grows. The density stays somewhat broad and only confined at low magnetic field ($B=1$ T), suggesting that the magnetic trapping is still constrained. The density, however, becomes tightly concentrated within the dot forming distinct resonant patterns when the field rises to intermediate and high values ($B=5.6$ T and $B=9.26$ T). This pattern shows how magnetic confinement causes quasibound states to form, whereby electrons execute repeated cyclotron motion and remain temporarily locked within the circular area \cite{Matulis2008}.
Though resonance peaks still show for particular magnetic fields, the density localization is weaker for the higher energy case ($E=1$ eV) than for the low-energy range. Because they have greater kinetic energy, higher-energy carriers are less susceptible to the restricting influence of the magnetic barrier and may readily exit the dot area. Less severely localized and more extended become the density profiles as a result. This agrees with the lowering of scattering efficiency seen in Figure~\ref{QB} for rising energy.
The presence of resonant magnetic confinement is shown by the emergence of strong localized density maxima at specific magnetic fields. These resonances result from the cyclotron orbit matching the effective size of the quantum dot, therefore generating constructive interference and intense carrier trapping. In graphene magnetic dots, where magnetic localization rather than electrostatic barriers causes confinement \cite{Matulis2008, Ramezani2010}, similar quasibound-state generation has been well investigated. WSe$_2$, unlike graphene, has a finite band gap and strong spin--orbit coupling, suppressing Klein tunneling and significantly improving localization efficiency \cite{Xiao2012}.
Because of the large Dirac spectrum and valley-dependent spin splitting, the density patterns in WSe$_2$ are more strongly localized and asymmetric, hence providing another significant contrast with graphene. Particularly at higher energies \cite{Matulis2008}, the gapless linear dispersion causes less confinement and wider density profiles in graphene. Conversely, the intrinsic mass element in WSe$_2$ increases the likelihood of confinement and generates sharper resonant states, hence increasing the magnetic quantum dot's ability for carrier capture.
Figure~\ref{RhoB} therefore affirms that the formation of quasibound states in WSe$_2$ quantum dots is controlled by the interaction between magnetic field intensity and carrier energy. Strong magnetic fields and low incident energies support highly localized resonant states, which are necessary for the development of spin- or valley-selective transport applications, tunable resonators, and quantum confinement devices.

\section{Conclusion}\label{con}

We have theoretically investigated the confinement and scattering properties of Dirac carriers in a circular quantum dot formed in monolayer WSe$_2$ under the presence of a perpendicular magnetic field. By solving the eigenvalue problem in the two spatial regions and imposing the continuity conditions at the interface of the magnetic quantum dot, we derived the corresponding wave functions and determined the scattering amplitudes associated with the different angular momentum channels.

Using these solutions, we analyzed the scattering efficiency as a function of the magnetic field strength, the quantum dot radius, and the incident carrier energy. Our results show that the total scattering efficiency is strongly enhanced at low incident energies and large magnetic fields, indicating efficient carrier confinement and the formation of quasibound states inside the dot. In particular, the lowest angular momentum modes, especially $l=0$ and $l=1$, were found to dominate the transport process, while higher-order channels remain weakly coupled due to stronger centrifugal suppression.

We also demonstrated that increasing the quantum dot radius significantly improves magnetic confinement by enlarging the effective interaction region and supporting stronger resonant states. The combined effect of large radius and strong magnetic field produces a substantial increase in scattering efficiency, confirming the cooperative role of geometric and magnetic confinement. Furthermore, the spatial density distributions reveal pronounced localization of the electronic states for specific magnetic fields, providing direct evidence of resonance formation and temporary carrier trapping inside the quantum dot.

Compared with graphene-based magnetic quantum dots, the confinement in WSe$_2$ is considerably stronger due to the presence of a finite band gap, strong spin--orbit coupling, and valley-dependent massive Dirac fermions, which suppress Klein tunneling and enhance localization. This makes WSe$_2$ a more efficient platform for magnetic confinement and quantum transport control.

Our findings demonstrate that magnetic quantum dots in WSe$_2$ provide an effective mechanism for tuning electron scattering, localization, and resonant transport. These results may be useful for the design of nanoscale quantum devices based on transition-metal dichalcogenides, including quantum resonators, spin filters, valley-selective transport systems, and other applications in spintronics and valleytronics.

\end{document}